# An Innovative Design of Substitution Box Using Trigonometric Transformation

Kashif Ishaq[1], Awais Ahmed Qarni[1]

[1]School of Systems and Technology, University of the Management and Technology, Lahore, Pakistan

**Corresponding Author:** kashif.ishaq@umt.edu.pk

**Abstract**

As the number of hacking events and cyber threats keeps going up, it's getting harder and harder to communicate securely and keep personal information safe on the Internet. Cryptography is a very important way to deal with these problems because it can secure data by changing it from one form to another. In this study, we show a new, lightweight algorithm that is based on trigonometric ideas and offers a lot of security by making it less likely that cryptanalysis will work. The performance of our suggested algorithm is better than that of older methods like the Hill cipher, Blowfish, and DES. Even though traditional methods offer good security, they may have more work to do, which slows them down. The suggested algorithm tries to close this gap by offering a solution based on trigonometric ideas that are both fast and safe. The main goal of this study is to come up with a small but strong encryption algorithm that can't be broken by cryptanalysis and keeps Internet communication safe. We want to speed up the coding process without making it less secure by using trigonometric principles. The suggested algorithm uses trigonometric functions and operations to create non-linearity and confusion, making it resistant to both differential and linear cryptanalysis. We show that the suggested algorithm is more secure and faster than traditional methods like the Hill cipher, Blowfish, and DES by doing a lot of research and testing. Combining trigonometric ideas with a simple design makes it workable for real-world uses and offers a promising way to protect data on the Internet.

## INTRODUCTION

Keeping and transmitting information in a way that only authorized recipients can read and route it is the goal of cryptography. Over the past six decades, the fundamentals of cryptography have been used in many different contexts. Governments or controlled groups have historically used it to conceal opposition opponents' covert communications. But at the same time, there are now daily millions of encrypted and private communications taking place online[1]. To protect data, photos, films, banking information, medical evidence, and much more, crypto methods are utilized. Because of this, there is a growing need for network security, as well as the science and research of ways to protect data in communication systems and computers from unauthorized changes. The daytime cryptography and network security practices are typically founded on mathematical theory and applications of computer science[2]. Data that is sensitive and confidential must thus be protected immediately given the explosive expansion of digital communication and data interchange. Symmetric encryption algorithms and asymmetric encryption algorithms are two categories of cryptography. Encryption techniques. The Advanced Encryption Standard (AES) and Data Encryption Standard are the two most popular. Symmetric encryption methods (DES). Symmetric Cryptography: The same key is utilized in this cryptographic system to both encrypt and decrypt the information processing. The key's length determines how reliable this method is. How to give the receiver the key is the problem with this method[3]. The security of the communication channel will likewise be managed if we send using the same communication cycle. Asymmetric Cryptography: Two keys are being utilized in this cryptographic mechanism. Tech individuals continue to deal with these innovations to make them secure and dependable for end clients, as information security. The sender of the information used his private key and public key for encoding. At the other end, the receiver used his private key and that public key for the process of decoding[4]. The issue of key distribution in symmetric cryptography is mitigated with this. Hashing: Sensitive information is translated into a fixed-length hash using the hash function. A better hash function must be used for more efficiency. It ensures that the data supplied must arrive exactly as it was sent, with no possibility of change. First, two categories are employed to ensure the privacy of the data. Hashing is used to guarantee the integrity of the data. Ciphers[5]. A series of operations called a cipher is used to encode and decode data using a unique key. A cipher is a name given to the combined operation of the encoding and decoding system. In other words, an algorithm was developed based on the symmetric and asymmetric ciphers that were developed. The Advance Encryption Standard and Data Encryption Standard are two well-known symmetric ciphers that have been applied in real-world situations[6]. Here are some of the most well- known asymmetric ciphers, including RSA, Diffie Hellman, and others. Here are some of the most common cipher types explained. Types of Ciphers For a better understanding of cipher processing, it's further divided into two categories based on the operation mechanism. Some algorithms operate whole input data in one go but some process by dividing the information's bit into multiple chunks and then processing each chunk separately. This working operation generates differences and comparisons in the reliability and efficiency of a cipher. Substitution Box The substitution box is nowadays one of the very vital components in the research field of cryptography and one of the hot topics these days. S-box works by the foundations of symmetric algorithms. As per the principles of Mathematics, it's a vectorial Boolean operation. S-box picks up an array of m bits and mutates it into an array of n bits. It's not a compulsory thing that an array of m is equal to that of n. The size of the m*n table is utilized as a lookup table to derive the answer. 17 S-box utilization in the field of Organizations and tech people working on these concepts to set some rules or procedures that not only make sure the confidentiality of data, integrity also not to be compromised.

Once organizations achieved such a set of procedures in counterpart, bad actors successfully managed to decode or get access to the data. Now the emerging demand is not only to make a set of procedures to secure data but also to secure that procedure so that bad actors cannot break it. Cryptography is an activity where a message or classified data written in a basic language transform into a language that is not justifiable. It is one of the procedures being involved these days and an arising point in the market to guarantee the security of information. Million-dollar subsidizing projects are being done to make cryptographic methods increasingly harder for assailants to break. Cryptography was first revealed before the world with the emergence of DES and AES-like algorithms, in which S-box is the major procedure[7] in their algorithms. It's a key-dependent formulation, which means to construct an S-box a variable-length key is required. The phenomenon of S-box working is the mutation of input data, which consists of a series of bits, into a new shaped output data. S-box can withstand the chances of compromise against both techniques of cryptanalysis i.e., linear and differential. Another important feature of the S-box is that it has attributes of both confusion and diffusion factor[8]. The strength factor of the s-box is directly proportional to its robustness. S-box is further categorized into two modes: Static S-box: Static S-box is a type of S-box that is pre-defined with the algorithm's procedure. That s-box is publicly available to everyone, and everyone can use it to encrypt or decrypt ciphertext. This makes the process weaker and an easy reverse mechanism for the attackers. So, such algorithms are not secure and cannot be utilized for implementation and there is no possibility of secure communication over the internet. Example of publicly available static boxes of DES, AES, etc. As in DES, an 8bit entry is fed into the s-box and with publicly available differential. Another important feature of the S-box is that it has attributes of both confusion and diffusion factor. The strength factor of the s-box is directly proportional to its robustness. S-box is further categorized into two modes: Static S-box: Static S-box is a type of S-box that is pre-defined with the algorithm's procedure. That s-box is publicly available to everyone, and everyone can use it to encrypt or decrypt ciphertext. This makes the process weaker and an easy reverse mechanism for the attackers. So, such algorithms are not secure and cannot be utilized for implementation and there is no possibility of secure communication over the internet. Example of publicly available static boxes of DES, AES, etc. As in DES, an 8-bit entry is fed into the s-box and with publicly available s-boxes responses are generated. Dynamix S-box: Needs raised to fix the deficiency of static S-box so the communication over the public channels will be secured. With the aid of dynamic s-box attackers or intruders don't have any idea how to generate an s-box which will help in the decryption phase. This 18 is one of the hottest topics in the field of cyber security and million-dollar researchers are performing it. Dynamic s-box is usually generated with the aid of a secret key. Without this key no one can generate that s-box, which is used to cipher the message, this makes the whole process secure. If a dynamic s-box generated with the use of a key for the AES algorithm it turns the security to 100NL (c) = 1 2 [2 (max 0,1 — ()—)] Algebraic Degree: This is another test to verify the s-box's security, the higher its degree and higher its complexity the better the S-box function will be. The algebraic normal form is one of the functions which depicts its polynomial value of it[9]. This algebraic degree is defined as the count of the coefficient that has a non-zero value. 19 Differential Uniformity: This test is performed on s-box on various pairs of s-box, the operation involved in this testing is XOR operation[10]. The result of the whole operation compiled multiple answers about each input value.

## BACKGROUND

S-box cryptography is a significant concept in current symmetric key cryptography, representing the heart of substitution- permutation networks (SPNs) used in popular block ciphers like the Advanced Encryption Standard (AES)[11]. The subsite- tuition box (S-box) is important to S-box cryptography. It is a mathematical entity or table that performs nonlinear replacements, infusing complexity into data changes. This complexity encourages the obfuscation of the connection be- tween input and output, increasing the complexity of cryptographic processes and enhancing security. Nonlinearity to prevent algebraic deductions, the avalanche effect to accent- tubate output discrepancies caused by minor input changes, and resistance to differential and linear cryptanalysis are all important characteristics of robust S-boxes. These features combine to constitute a cornerstone in the design of robust encryption algorithms, as demonstrated by the use of S- boxes inside them[12]. Mathematicians refer to the complex and erratic behaviors that nonlinear dynamical systems display as chaos. These behaviors are distinguished by their sensitivity to the beginning conditions, leading to erroneous oscillations, bifurcations, and the appearance of odd attractors. Chaotic systems are useful in many fields, including physics, biology, chemistry, engineering, and economics because they are good models for a variety of phenomena. Notably, secure common- nidation, data encryption, and signal processing are among the practical applications of chaotic systems. Due to their fundamental structure and their importance in cryptography, one-dimensional (1D)[12] chaotic systems have attracted a lot of attention. However, it has been shown that the majority of 1D chaotic systems have a limited spectrum of chaos. Given this restriction, investigating compound chaotic maps becomes an appealing option. Particularly in the field of cryptography, these compound maps have the singular capacity to provide enhanced chaotic behaviors and increased security in a variety of applications[13]. Due to their fundamental structure and their importance in cryptography, one-dimensional (1D) chaotic systems have attracted a lot of attention. However, it has been shown that the majority of 1D chaotic systems have a limited spectrum of chaos. Given this restriction, investigating com- pound chaotic maps become an appealing option. Particularly in the field of cryptography, these compound maps have the singular capacity to provide enhanced chaotic behaviors and increased security in a variety of applications.

**RELATED WORK**

A substantial amount of research is being carried out in the cryptography domain; one of the hot topics is the substitution box (S-box). Bukhari et al[14]. proposed a novel model to encrypt grayscale images using S-box. The proposed approach uses the Galois field and linear fractional transformations to generate six different S-boxes and then use them for gray-scale image encryption. Shah and Shah [15] developed a new method for creating sixteen (16) distinct robust $8 \times 8$ S-boxes using sixteen extensions of the Galois field GF(28 ). In the proposed methodology, sixteen linear fractional transformations were defined against those Galois fields. To improve the encryption strength of AES, Hussain et al[15].Employed the chaotic logistic map, Mobius transformation, and symmetric group S256 to create a dynamic S-box. Shah et al. [16] proposed a 24× 24 S-box. For generating the proposed S-box, the maximal cyclic subgroup of the multiplicative group of Galois ring unit GR (23, 8) was used. The proposeS-box has better confusion potential than that of any $8 \times 8$ S-boxes. The authors used the proposed S-box in encryption to induce confusion in RGB channels of a plain image. $24 \times 24$ S-box dependent encryption method outperformed the $8 \times 8$ S-box dependent encryption method. Gao et al. [15] presented a technique for generating S-box utilizing a module of a group of PSL(2, Z) on projective line PL(F257) over Galois Field GF (28 ). The obtained results were compared with relevant S-Boxes. Chew and Ismail [17] presented a new way to design and construct cryptographically strong $8 \times 8$ S-boxes for block ciphers utilizing a linear fractional transformation and a permutation function over the GF(28 ).

**PROPOSED APPROACH FOR S-BOX DESIGN**

Those block ciphers designed via chaotic maps show more. On-linearity and efficiency when compared with commonly known ciphers. LMs based on CM as one of the fields of mathematics require complex mathematical calculations to design a formula for encryption[17]. The formula prepared for this thesis work is based on CM, a key-dependent mechanism developed to prepare a dynamic S-box. The chaos property of CM is widely used to develop entropy, this feature generates the properties of confusion and diffusion[18]. When the chaotic parameters along with the encryption key are merged system- atically, it becomes almost impossible for attackers to predict the key or plain text from the output[19]. S-box Construction Process In this Paperwork, a unique formula was devised to produce a dynamic S-box. The following equation was devised for its preparation:

$$Xn + 1 = \begin{cases} 0.52 + Xn + Sin(A * Xn) * Sin(A * Xn), & x < 0.5 \\ 1.5f + \dfrac{A}{2.0} - Tan(Xn), & otherwise \end{cases}$$

## BIFURCATION

The study of bifurcation includes both topological and qualitative aspects. The phase space of a system changes as a ret of. Has significant threshold variations and parameters. Solid principles. Are shown by a solid line, while an unstable area is shown by a .Ted line. Values. Usually, a small change in the parameters is the cause. A significant change in phase spat affects system performance[20].

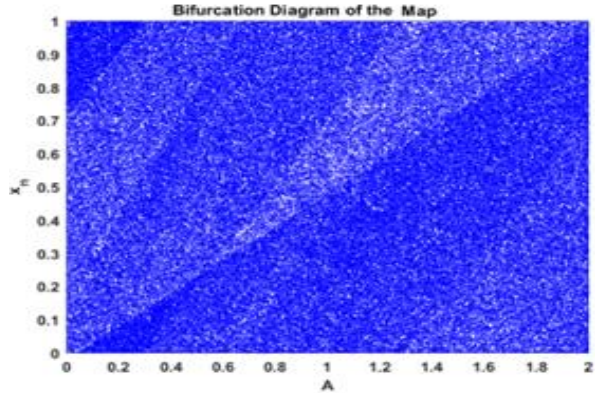

## LYAPUNOV EXPONENT

For the purpose of describing chaos, an analytical metric known as the Lyapunov-Exponent (LE) is utilized. The system is chaotic if the chaotic map of the system has a Lyapunov exponent that is greater than 0. The instability of the system increases in proportion to the value of LE. The rate at which the unseen approximation is either converging or diverging is disclosed by the LE. In cryptography, the idea of Lyapunov exponents can also be utilized for the purpose of performing analysis on substitution boxes, commonly known as S-boxes. S-boxes, which are essential elements of symmetric key block ciphers, are the ones accountable for the non-linearity and confusion that are introduced throughout the encryption process. By gaining an understanding of the Lyapunov exponents of S-boxes, one can gain insights into the dynamical behavior and security properties of these structures.In order to use the Lyapunov exponents with S-boxes, we must first think of the S-box as a discrete dynamical system. The input to the S-box is a binary vector, and the output is also a binary vector that is created by applying the S-box transformation. Both vectors can be expressed using the same notation. After that, we will be able to investigate how slight alterations in the input are transmitted across the S-box and impact the output[21].

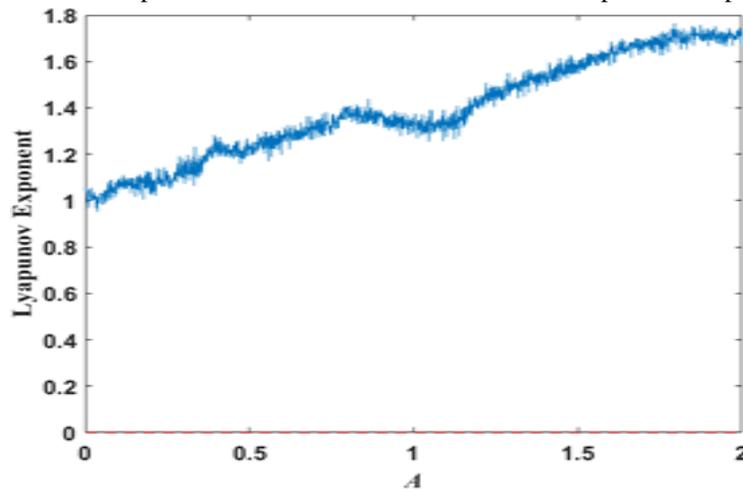

```
Algorithm 01 Initial S-Box
Input Values:
    X             // 0 < X < 1.0
    A             // 0 < A < 2.0
    B             // 0 < A < 10^9

Output values:
    S, B          // Arrays of size 256 each
Initializations:
    Loc ← 0
Procedure:
    WHILE (Loc <= 255) DO
        W ← 1.0 - Cos(Xn) + (3+ A) / Xn
        Y ← Sqrt( Sin(Xn)/0.5 + A * Xn
        X ← MOD ( Round( Xn * B, 0) , 256)

        S[Loc] ← X

        Loc ← Loc + 1
    END WHILE
```

## PROPOSED S-BOX SECURITY ANALYSIS

A significant data and information research contribution. The development of new S-boxes is central to the security field. After an S-box is created, it is examined to determine its. The ability to choose how strong it will be against various attacks. (Differential and linear). There are tests for an S-Box's cryptanalytic evaluation. Calculated using the predetermined criteria, which include:

a. The Bijective Ness.
b. The Nonlinearity.
c. The Fixed Points.
d. The Strict Avalanche Criterion
e. The Bit Independence Criterion.
f. The Linear Approximation Probability.
g. The Differential Approximation Probability.

### a. The Bijective Ness

The study of bifurcation includes both topological and qualitative aspects. The phase space of a system changes as a set of. There are critical threshold variations and parameters. solid principles. is indicated by a solid line and unstable regions are indicated by black. Ted line. Values. Usually, a small change in the parameters is the cause. A significant change in phase space affects system performance[21].

### b. The Nonlinearity.

The result of the Non-Linearity (NL) of the proposed s-box is shown in Table 3. Then a comparison of the proposed S- box's NL is performed with other s-boxes as shown in Table 3 and Figure 4.

| NL Value | 106 | 106 | 106 | 106 | 106 | 108 | 106 | 108 |
|---|---|---|---|---|---|---|---|---|

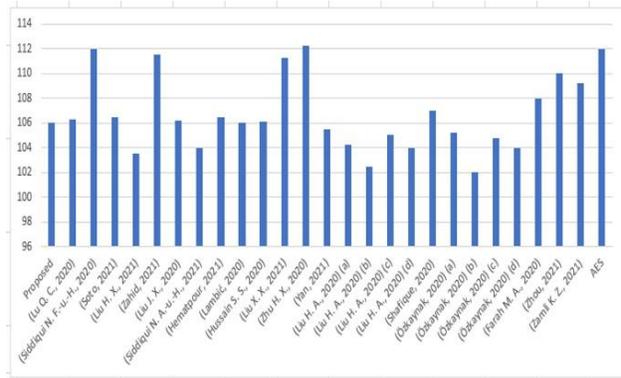

### c. The Fixed Points

If there is any fixed point in a substitution box, an attacker may be able to decipher the secret data from the captured ciphertext. For safety reasons, the proposed S-Box was tested according to fixed point criteria. Therefore, there may be no fixed points in the final S-Box[22].

### d. The Strict Avalanche Criterion

```
0.4375  0.5156  0.5313  0.4531  0.4531  0.5313  0.5313  0.5000
0.5156  0.4219  0.5000  0.4688  0.4688  0.4844  0.5156  0.5313
0.5469  0.5469  0.4688  0.5313  0.5469  0.5469  0.4531  0.5313
0.5156  0.5313  0.5938  0.4844  0.5313  0.5156  0.5313  0.4531
0.4531  0.5469  0.5000  0.5000  0.5000  0.5469  0.5469  0.5156
0.4531  0.5156  0.5156  0.5000  0.5156  0.5000  0.5781  0.5469
0.4688  0.5625  0.5469  0.5000  0.5625  0.4531  0.5156  0.4375
0.4844  0.4375  0.4531  0.4531  0.5313  0.5938  0.4531  0.5469
```

The SAC addresses the sensitivity to the slightest alteration in the input data. Tavares and Webster described the concept of SAC based on the complementary effect and the avalanche [55]. This criterion gauges how many output bits are altered by a single change to the input When only one input bit is flipped, SAC is said to be satisfied when all output bits flip with a probability of 0.5. The table displays the calculated SAC scores for each of the four S-boxes. As a result of their values being relatively close to the ideal value, the results demonstrate that the suggested S-boxes satisfy the strict avalanche criteria[23].

### e. The Bit Independence Criterion.

An additional criterion for judging S-Box performance was developed by Tavares and Webster and is called the Bit Independence Criterion (BIC). These standard states that if the input bits change, the output bits must also change. The SAC and BIC-NL values of several S-Boxes are compared. Searches are conducted using the independence of bits criterion on input bits that remain constant. The advantages of this technique include independent performance optimization of variables using pairs of unmodified input bits and avalanche vectors. In symmetric cryptosystems, this is a useful standard due to the improved independence of the bits[24].

f. **The Linear Approximation Probability.**

The linear approximation probability (LAP) method is use-full for determining an incident's imbalance. Linear Probability is also one of the tests to prove the competence of the S- box. The relation between the LP and the strength of the S- box is inverse. It means lessened the value of LP the higher the strength of the S-box. The robust feature of the S-box eventually raised with the lesser value of the LP. The LP value of the proposed S-box is 0.1406.

```
0.0000 0.5273 0.5156 0.4980 0.5039 0.5176 0.5234 0.5000
0.5273 0.0000 0.5000 0.5137 0.4727 0.4980 0.5215 0.5020
0.5156 0.5000 0.0000 0.4844 0.5020 0.4844 0.5078 0.4922
0.4980 0.5137 0.4844 0.0000 0.5176 0.5000 0.5254 0.4883
0.5039 0.4727 0.5020 0.5176 0.0000 0.5156 0.5000 0.5254
0.5176 0.4980 0.4844 0.5000 0.5156 0.0000 0.5156 0.5117
0.5234 0.5215 0.5078 0.5254 0.5000 0.5156 0.0000 0.5195
0.5000 0.5020 0.4922 0.4883 0.5254 0.5117 0.5195 0.0000
```

Fig. 7. Average BIC-SAC Value = 0.5066

g. **The Differential Approximation Probability**.

The linear approximation probability (LAP) method is use-full for determining an incident's imbalance. Linear Probability is also one of the tests to prove the competence of the S- box. The

| 8 | 8 | 6 | 10 | 8 | 6 | 8 | 8 | 6 | 6 | 8 | 6 | 8 | 8 | 6 | 6 |
|---|---|---|---|---|---|---|---|---|---|---|---|---|---|---|---|
| 6 | 6 | 8 | 6 | 8 | 8 | 8 | 6 | 6 | 6 | 6 | 8 | 10 | 8 | 6 | 6 |
| 6 | 8 | 8 | 6 | 6 | 6 | 6 | 8 | 6 | 6 | 6 | 6 | 8 | 6 | 6 | 6 |
| 6 | 8 | 6 | 6 | 6 | 6 | 8 | 6 | 6 | 10 | 6 | 6 | 8 | 8 | 8 | 6 |
| 6 | 6 | 6 | 8 | 6 | 8 | 8 | 6 | 8 | 8 | 8 | 6 | 8 | 6 | 6 | 8 |
| 6 | 6 | 8 | 8 | 8 | 6 | 8 | 6 | 6 | 8 | 8 | 6 | 8 | 6 | 8 | 8 |
| 6 | 6 | 6 | 6 | 6 | 6 | 6 | 6 | 8 | 6 | 6 | 8 | 4 | 6 | 8 | 6 |
| 6 | 6 | 6 | 6 | 8 | 6 | 6 | 8 | 8 | 8 | 6 | 6 | 8 | 4 | 6 | 8 |
| 8 | 6 | 6 | 8 | 6 | 10 | 8 | 8 | 6 | 6 | 6 | 6 | 6 | 8 | 6 | 6 |
| 6 | 6 | 6 | 8 | 6 | 6 | 8 | 8 | 10 | 8 | 10 | 8 | 10 | 6 | 8 | 10 |
| 6 | 6 | 8 | 8 | 6 | 6 | 8 | 8 | 8 | 8 | 8 | 6 | 6 | 6 | 8 | 10 |
| 8 | 6 | 6 | 8 | 6 | 6 | 8 | 8 | 6 | 6 | 8 | 6 | 6 | 8 | 8 | 6 |
| 6 | 6 | 6 | 6 | 6 | 8 | 10 | 8 | 6 | 4 | 8 | 6 | 8 | 8 | 6 | 8 |
| 6 | 6 | 6 | 6 | 6 | 6 | 6 | 6 | 6 | 6 | 6 | 6 | 8 | 8 | 6 | 6 |
| 6 | 6 | 8 | 8 | 6 | 8 | 6 | 6 | 8 | 6 | 6 | 10 | 8 | 6 | 6 | 6 |
| 6 | 6 | 6 | 8 | 6 | 4 | 6 | 6 | 6 | 8 | 8 | 10 | 6 | 6 | 8 | 0 |

relation between the LP and the strength of the S- box is inverse. It means lessened the value of LP the higher the strength of the S-box. The robust feature of the S-box eventually raised with the lesser value of the LP. The LP value of the proposed S-box is 0.1406[24].

## FLOWCHART INITIAL AND FINAL S-BOX OF PROPOSED S-BOX METHOD

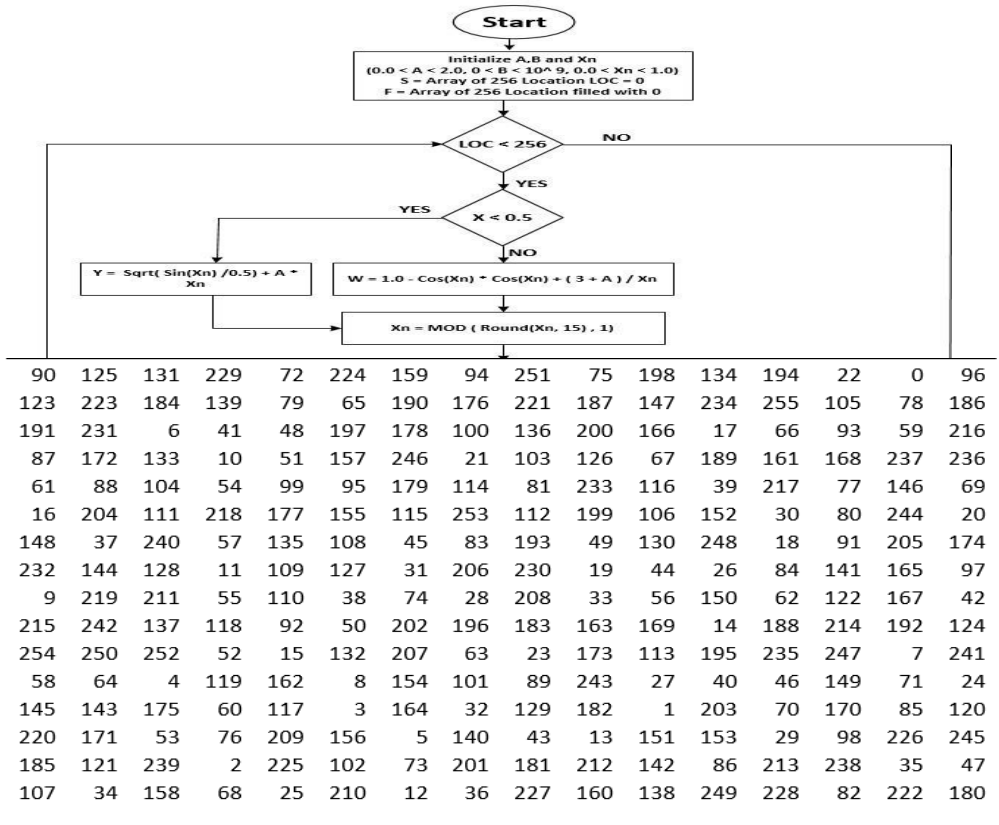

| 90 | 125 | 131 | 229 | 72 | 224 | 159 | 94 | 251 | 75 | 198 | 134 | 194 | 22 | 0 | 96 |
|---|---|---|---|---|---|---|---|---|---|---|---|---|---|---|---|
| 123 | 223 | 184 | 139 | 79 | 65 | 190 | 176 | 221 | 187 | 147 | 234 | 255 | 105 | 78 | 186 |
| 191 | 231 | 6 | 41 | 48 | 197 | 178 | 100 | 136 | 200 | 166 | 17 | 66 | 93 | 59 | 216 |
| 87 | 172 | 133 | 10 | 51 | 157 | 246 | 21 | 103 | 126 | 67 | 189 | 161 | 168 | 237 | 236 |
| 61 | 88 | 104 | 54 | 99 | 95 | 179 | 114 | 81 | 233 | 116 | 39 | 217 | 77 | 146 | 69 |
| 16 | 204 | 111 | 218 | 177 | 155 | 115 | 253 | 112 | 199 | 106 | 152 | 30 | 80 | 244 | 20 |
| 148 | 37 | 240 | 57 | 135 | 108 | 45 | 83 | 193 | 49 | 130 | 248 | 18 | 91 | 205 | 174 |
| 232 | 144 | 128 | 11 | 109 | 127 | 31 | 206 | 230 | 19 | 44 | 26 | 84 | 141 | 165 | 97 |
| 9 | 219 | 211 | 55 | 110 | 38 | 74 | 28 | 208 | 33 | 56 | 150 | 62 | 122 | 167 | 42 |
| 215 | 242 | 137 | 118 | 92 | 50 | 202 | 196 | 183 | 163 | 169 | 14 | 188 | 214 | 192 | 124 |
| 254 | 250 | 252 | 52 | 15 | 132 | 207 | 63 | 23 | 173 | 113 | 195 | 235 | 247 | 7 | 241 |
| 58 | 64 | 4 | 119 | 162 | 8 | 154 | 101 | 89 | 243 | 27 | 40 | 46 | 149 | 71 | 24 |
| 145 | 143 | 175 | 60 | 117 | 3 | 164 | 32 | 129 | 182 | 1 | 203 | 70 | 170 | 85 | 120 |
| 220 | 171 | 53 | 76 | 209 | 156 | 5 | 140 | 43 | 13 | 151 | 153 | 29 | 98 | 226 | 245 |
| 185 | 121 | 239 | 2 | 225 | 102 | 73 | 201 | 181 | 212 | 142 | 86 | 213 | 238 | 35 | 47 |
| 107 | 34 | 158 | 68 | 25 | 210 | 12 | 36 | 227 | 160 | 138 | 249 | 228 | 82 | 222 | 180 |

Initial S-Box

| 90 | 125 | 131 | 229 | 72 | 224 | 159 | 94 | 251 | 75 | 198 | 134 | 194 | 22 | 0 | 96 |
|---|---|---|---|---|---|---|---|---|---|---|---|---|---|---|---|
| 123 | 223 | 184 | 139 | 79 | 65 | 190 | 176 | 221 | 187 | 147 | 234 | 255 | 105 | 78 | 186 |
| 191 | 231 | 6 | 41 | 48 | 197 | 178 | 100 | 136 | 200 | 166 | 17 | 66 | 93 | 59 | 216 |
| 87 | 172 | 133 | 10 | 51 | 157 | 246 | 21 | 103 | 126 | 67 | 189 | 161 | 168 | 237 | 236 |
| 61 | 88 | 104 | 54 | 99 | 95 | 179 | 114 | 81 | 233 | 116 | 39 | 217 | 77 | 146 | 69 |
| 16 | 204 | 111 | 218 | 177 | 155 | 115 | 253 | 112 | 199 | 106 | 152 | 30 | 80 | 244 | 20 |
| 148 | 37 | 240 | 57 | 135 | 108 | 45 | 83 | 193 | 49 | 130 | 248 | 18 | 91 | 205 | 174 |
| 232 | 144 | 128 | 11 | 109 | 127 | 31 | 206 | 230 | 19 | 44 | 26 | 84 | 141 | 165 | 97 |
| 9 | 219 | 211 | 55 | 110 | 38 | 74 | 28 | 208 | 33 | 56 | 150 | 62 | 122 | 167 | 42 |
| 215 | 242 | 137 | 118 | 92 | 50 | 202 | 196 | 183 | 163 | 169 | 14 | 188 | 214 | 192 | 124 |
| 254 | 250 | 252 | 52 | 15 | 132 | 207 | 63 | 23 | 173 | 113 | 195 | 235 | 247 | 7 | 241 |
| 58 | 64 | 4 | 119 | 162 | 8 | 154 | 101 | 89 | 243 | 27 | 40 | 46 | 149 | 71 | 24 |
| 145 | 143 | 175 | 60 | 117 | 3 | 164 | 32 | 129 | 182 | 1 | 203 | 70 | 170 | 85 | 120 |
| 220 | 171 | 53 | 76 | 209 | 156 | 5 | 140 | 43 | 13 | 151 | 153 | 29 | 98 | 226 | 245 |
| 185 | 121 | 239 | 2 | 225 | 102 | 73 | 201 | 181 | 212 | 142 | 86 | 213 | 238 | 35 | 47 |
| 107 | 34 | 158 | 68 | 25 | 210 | 12 | 36 | 227 | 160 | 138 | 249 | 228 | 82 | 222 | 180 |

Final S-Box

# EFFICIENCY ANALYSIS

On a machine running Windows 7 with 6GB of Memory and a 2.24GHz Intel Core i5 Processor, to assess the computational efficiency of the suggested S-box approach, a Visual C simulation was done. The suggested method's calculation efficiency was observed for both the initial and final S-boxes. A clever and heuristic approach for calculating the initial S- cryptographic box's strength is necessary for the S-creation. box's 100000 distinct beginning S-boxes were produced in order to assess their time complexity and the time needed to the average time complexity of these initial and final S-box constructions is measured in Table 10. Table 10 demonstrates how highly motivating the preliminary S-building box's time is. However, the proposed solution requires a little more time to build an S-box.

# CONCLUSION

There are two popular design approaches for creating S- boxes: chaos-based and Trigonometric Transformation theory- based. The security benefits of each of these design paradigms are distinct. application. In this article, we compare the two approaches and propose a way to construct a non-bijective S-box. The proposed improved chaos map of this methodis used for the chaotic heuristic search of the first S-box. The security properties of the generated bijective S-boxes are improved by a proposed strong group created after extensive experiments. Algebraic group actions improvise cryptographic properties of S-boxes. Based on experimental results. findings,the suggested S-boxes adequately met the requirements for competent S-boxes. The ability to build highly nonlinear, Variable-sized S-boxes are an advantage of the proposed S-boxgeneration method. Another advantage of this method is that ityields the highest nonlinearity score of 109.5 for an S-box of any of the available S-boxes. Additionally, S-boxes of recent proposals are used in the comparative analysis to determine theoverall standing. Comparing the proposed bijective S-boxes tothe majority of the existing S-boxes, it has been discovered thatthey have superior nonlinearity, SAC, and linear approximationprobability features[24]